\documentclass{aa}
\usepackage{graphicx,times}
\newlength{\fsize}
\newcommand{\mk}{}

\def\Real{{\rm I\mathchoice{\kern-0.70mm}{\kern-0.70mm}{\kern-0.65mm}%
  {\kern-0.50mm}R}}  
  \def\bx#1{\leavevmode\thinspace\hbox{\vrule\vtop{\vbox{\hrule\kern1pt
  \hbox{\vphantom{\tt/}\thinspace{\bf#1}\thinspace}}
  \kern1pt\hrule}\vrule}\thinspace}

\def\be{\begin{equation}} \def\ee{\end{equation}}

\begin{document}

\title{The stability of toroidal fields in stars}

\author{J. Braithwaite}
\institute{Max-Planck-Institut f\"ur Astrophysik, Karl-Schwarzschild-Stra{\ss}e 1,
Postfach 1317,  D--85741 Garching, Germany\\
\tt jon@mpa-garching.mpg.de}
\authorrunning{J. Braithwaite}
\titlerunning{The stability of toroidal fields in stars}

\abstract{We present numerical models of hydromagnetic instabilities under the
conditions prevailing in a stably stratified, non-convective stellar interior, 
and compare them with previous results of analytic work on instabilities in 
purely toroidal fields. We 
confirm that an $m=1$ mode (`kink') is  the dominant instability in a toroidal 
field in which the field strength is proportional to distance from the axis, such
as the field formed by the winding up of a weak field by differential rotation. 
We measure the growth rate of the instability as a function of field strength and
rotation rate $\Omega$, and investigate the effects of a stabilising thermal 
stratification as well as magnetic and thermal diffusion on the stability. Where
comparison is computationally feasible, the results agree with analytic 
predictions. 
\keywords {instabilities -- magnetohydrodynamics (MHD) -- stars:
magnetic fields}}

\maketitle

\section{Introduction}

Magnetic fields probably play a significant role in the internal
rotation of stars. Even a relatively weak magnetic field is sufficient
to couple different parts of the star and maintain a state of nearly
uniform rotation. For the interior of the present Sun, for example, a
field of less than 1 gauss would be able to transmit the torque
exerted by the solar wind through the interior (\cite{Mestel:1953}). The
observed rotation in the core of the Sun (\cite{Chaplinetal:2001}) is quite uniform,
suggesting that a magnetic field of this order or larger may actually
be present. The progenitors of white dwarfs and supernovae go
through giant stages in which the envelope rotates very slowly. The
degree of coupling between core and envelope by a magnetic field in
this stage will determine whether the rotation rates of pulsars and
white dwarfs are just a remnant of the initial rotation of their
progenitors, or if a secondary process must be responsible
(\cite{SprandPhi:1998}, \cite{Spruit:1998dn}).

Models of gamma-ray bursts in which the
central engine derives from the rapidly spinning core of a massive
star (\cite{Woosley:1993}, \cite{Paczynski:1998}) also depend on the ability of the core
to keep its high angular momentum for a sufficient period of time, in
the face of magnetic spindown torques exerted by the slowly rotating
envelope (\cite{Hegeretal:2000}).

The uniform rotation of the solar core may be due to a magnetic field,
but this field's origin, configuration and strength are not known. By analogy
with the magnetic A-stars, one might speculate that a `fossil'
magnetic field could exist in the core of the Sun {\mk(\cite{Cowling:1945}, \cite{BraandSpr:2004}).} Since no net field
is seen at the surface (averaged over the solar cycle), the radial
component of such a fossil would however have to be weak -- of the
order of a gauss or less. A field weaker than about this 1 G will
quickly wrap up into a predominantly toroidal field, under the action
of the remaining (weak) differential rotation in the core.

The predominantly toroidal magnetic field resulting from this process
will not increase in strength arbitrarily. Eventually, the energy density
in the field will become large enough that a magnetic instability will set
in.

Analytic work, (e.g. \cite{Tayler:1973}), shows that any purely toroidal
field should be unstable to instabilities on the magnetic axis of the
star (pinch-type instabilities, under the influence of the strongly
stabilising stratification in a radiative stellar interior, or `Tayler
instabilities' hereafter). The growth rates of these instabilities are
expected to be of the order of the time taken for an Alfv\'{e}n wave
to travel around the star on a toroidal field line. This is very
short compared to the evolutionary timescale of the star. 
In a star like the Sun, for example, with a field of $1000$ gauss, 
the growth timescale $r\sqrt{4\pi\rho}/B$ would be of the order 
of years, if $r=R_\odot/2$ is taken, and 
$\rho=1.3 \mathrm{g/cm}^3$.

{\mk A magnetic field of this type
can also be subject to other instabilities, such as the magnetic
buoyancy (\cite{Parker:1955}, \cite{Gilman:1970}, \cite{Acheson:1978})
and magnetic shear instabilities (\cite{Velikhov:1959},
\cite{Acheson:1978}, \cite{BalandHaw:1992}). As was
shown by Spruit (1999), the Tayler instability will be the first to
appear as the strength of the toroidal field is increased. This is
because with the
magnetic buoyancy instability, as with all instabilities where
displacements in the vertical are necessary, the stratification
provides a strong stabilising force. The same is the case with the
magnetic shear instability, whose effect in a stellar interior is very
limited in comparison to its effect in accretion discs. In contrast, the Tayler
instability occurs on the magnetic axis, where the
magnetic field is perpendicular to gravity and the displacements
caused by the instability are also perpendicular to gravity.}

In this paper we aim to test numerically the instability mechanism,
and to verify that the predictions of the analytic work are relevant:
that they cover all instabilities actually present in a system
consisting of a predominantly toroidal field in a stable
stratification. Much of the analytical stability analyses have been
done under a local approximation. This can be shown to be exact for
the case of adiabatic instabilities in a non-rotating star, but not
for the more interesting cases in which rotation and the effects
of magnetic and thermal diffusion are taken into account. Though 
it is not expected that major instabilities have been missed, numerical 
simulations can provide an important check.

It is much less certain how the magnetic field evolves once
instability has set in. In a scenario developed by Spruit (2002), it is argued that
the instability will lead to self-sustained dynamo action. The
field remains predominantly toroidal, subject to decay by 
Tayler instability, but is continuously regenerated by the winding-up of 
irregularities produced by the instability. This scenario has been
applied in stellar evolution calculations of the internal rotation of massive 
stars by Heger et al. (2003) and Maeder and Meynet (2003). 

The balance between wrapping-up by differential rotation on the one
hand and the destruction of the toroidal field by Tayler instabilities
on the other determines the strength and configuration of the field,
and the the rate at which it transports angular momentum through the
star.

{\mk The long-term goal is therefore to investigate the non-linear development of the
instability. With 3-D numerical simulations we can determine how quickly
an initial toroidal field decays (by reconnection across the magnetic
axis), and to determine the type of magnetic field that is
maintained by differential rotation in a stably stratified star, under
the action of magnetic instabilities, and to develop from this a
quantitative theory for the angular momentum transport by magnetic
fields in stars. This is beyond the scope of this paper, but is looked
at by Braithwaite (2005), where the operation of this
differential-rotation driven magnetic dynamo is demonstrated.}

%

\section{The nature of the instabilities}

We first consider adiabatic instabilities, that is, ignoring the
effects of viscosity and of thermal and magnetic diffusion. The
instabilities of a magnetic field in a stable stratification then
depend on three parameters: the field strength, some measure of the
stability of the stratification, and the rotation rate of the
star. For stars rotating well below their critical (maximal) rate, and
for the expected relatively low field strengths, the relative strength
of the parameters is expressed by the ordering

\begin{figure}
\includegraphics[width=1.0\hsize,angle=0]{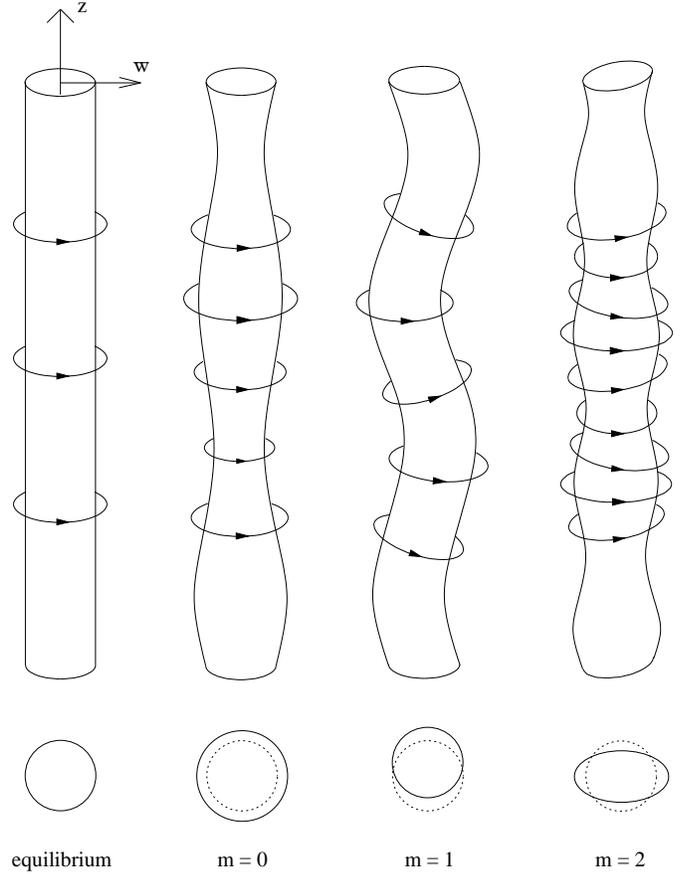}
\caption{The physical form of the instability, for modes $m=0$, $m=1$
and $m=2$. Above, co-moving surfaces are drawn, with the magnetic
field represented by the lines with arrows. {\mk The $z$ and $\varpi$ axes
are marked on the diagram on the left; gravity (if present)
acts along the $z$ axis.} Below, cross-sections of
these co-moving surfaces are plotted, with the dotted circles
representing equilibrium.}
\label{fig:fig1}
\end{figure}

\begin{equation}
N \gg \Omega\qquad\mbox{and}\qquad N \gg \Omega_A,
\end{equation}
where $N$ is the buoyancy frequency, $\Omega$ is the rotational
velocity, and $\Omega_{\rm A}$ is the Alfv\'{e}n frequency given by
$v_{\rm A}/R=B/(R \sqrt{4\pi \rho})$, $R$ is the radius of the star and
$v_{\rm A}$ is the Alfv\'{e}n speed. This is the
same, in effect, as saying that the thermal energy density is much
greater than both the rotational and the magnetic energy densities.

We consider an azimuthal field $\mathbf{B} = \mathbf{e}_\phi
B(\varpi)$. We define
\begin{equation}
p\equiv \frac{\partial \ln B}{\partial \ln \varpi} \qquad\mbox{as}\qquad \varpi
\rightarrow 0.
\end{equation}

In cylindrical coordinates $(\varpi,\phi,z)$ where the $z$ axis is
parallel to the magnetic axis, the
displacements from equilibrium are of the form
\begin{equation}
\xi \sim e^{ i(l\varpi + m\phi + nz) + \sigma t}.
\end{equation}

The shape of the unstable displacements is shown in Fig. \ref{fig:fig1},
for modes with $m=0,1,2$. {\mk The $m=1$ mode is known as the `kink'
instability, and is seen
in twisted solar coronal loops and twisted rubber bands. This
instability sets in once the magnetic field
component along the flux tube (vertical in the geometry used here)
falls below some critical value in relation to the toroidal component. In
our case of a predominantly toroidal field, resulting from winding up
by differential rotation, the vertical component is neglected: it is a limiting case.}

It is predicted that for an instability to occur, we must have (\cite{Tayler:1957})
\begin{equation}
p > \frac{m^2}{2} - 1 \quad (m\neq 0) \qquad\mbox{and}\qquad p > 1
\quad (m=0).
\label{eq:mandp}
\end{equation}

{\mk It is now useful to consider the likely value of $p$ in a
differentially-rotating star. The toroidal field wound up by differential rotation will be:
\begin{equation}
\frac{\partial B_\phi}{\partial t} =
B_{\rm z}^{\rm init}\varpi \frac{\partial\Omega}{\partial z} + 
B_{\rm \varpi}^{\rm init}\varpi \frac{\partial\Omega}{\partial \varpi},
\label{eq:b_dep_on_varpi}
\end{equation}
where $B_{\rm z}^{\rm init}$ and $B_{\rm \varpi}^{\rm init}$ are the
initial field components.
We expect both spatial derivatives of $\Omega$ to tend to some finite
value on the z-axis, whether the differential rotation is being driven
by braking on the surface or by evolutionary changes in the moment of
inertia, etc. There is no obvious reason why $B_{\rm z}^{\rm init}$
should go to zero on the axis, so the first term on the right-hand
side of
Eq.~\ref{eq:b_dep_on_varpi} will be proportional to $\varpi$. The
behaviour of $B_{\rm \varpi}^{\rm init}$ near the z-axis is less
certain, and it may have some dependence on $\varpi$ so that the
second term goes as $\varpi^2$ or higher, but overall the toroidal field
component $B_\phi$ will be proportional to $\varpi$, i.e. $p=1$.

So if we use $p=1$, the} only unstable mode is $m=1$; $m=0$ and $m=2$ are
marginally stable. If $p=2$, then we should expect the $m=0$, $m=1$
and $m=2$ modes all to be unstable.

Only high values of the vertical wavenumber $n$ are unstable. This is due to the
dominant effect of the stable stratification. In order to minimise the energy lost against the stable
buoyancy force, the vertical displacements have to be small compared to the horizontal displacement. Since the displacements also have to be nearly incompressive (otherwise energy
is lost in doing work against the pressure force) this implies that the vertical length scale is small.

If we define a local Alfv\'{e}n frequency
\begin{equation}
\omega_{\rm A} \equiv v_{\rm A}/\varpi = B/(\varpi \sqrt{4\pi \rho}),
\label{eq:defofoa}
\end{equation}
then we expect the growth rate for a
non-rotating star to be (\cite{Tayler:1973} and \cite{Gooetal:1981})
\begin{equation}
\sigma \sim \omega_{\rm A} \quad\mbox{($\Omega\ll\omega_{\rm A}$)}.
\label{eq:growthratelo}
\end{equation}

The instability condition is local (in a meridional plane): if it is satisfied at 
some point ($\varpi,z$), an unstable eigenfunction can be fit into a small
(for adiabatic instability: infinitesimal) region around this point
(Tayler 1973). The instability is thus of an `interchange' type. This
property holds by virtue of the assumption of a purely toroidal
field. It greatly simplifies the analytical stability analysis, and
allows a detailed treatment of the effects of diffusion and
viscosity (\cite{Acheson:1978}). 
In the azimuthal direction, the instability is global, since only low 
azimuthal orders $m$ are unstable. Connected with this is the fact 
that the typical instability time scale (if conditions for instability are
satisfied) are always of the order of the time it takes an Alfv\'en
wave to travel around the star in the azimuthal direction.

If the star is rotating with the rotation axis parallel to the
magnetic axis, a stabilising effect is produced so that instead of
(\ref{eq:mandp}) we have (\cite{PitandTay:1986})

\begin{equation}
p > \frac{m^2}{2} + 1 \quad (m\neq 0) \qquad\mbox{and}\qquad p > 1
\quad (m=0).
\label{eq:mandprot}
\end{equation}

in the limiting case where the rotational velocity is much greater
than the magnetic. In the particular case of $p=1$, we expect
stability at all values of $m$. 
[We note that this is a peculiarity of the adiabatic case: when magnetic
and thermal diffusion are included, the conditions on $p$ for instability
reverts to (\ref{eq:mandp}) (cf. Appendix in Spruit 1999 and below).]

\subsection{The effect of magnetic diffusion}

As mentioned, there is a lower limit on the vertical wavenumber of an 
unstable mode, caused by the work which must be done against gravity to
move gas in the vertical direction. There is also an upper limit to
$n$, caused by diffusion: the unstable perturbations diffuse away at a
rate of order $\eta n^2$, so an instability whose intrinsic (adiabatic)
growth rate is less than its decay rate by diffusion will be
smothered. The range of unstable radial wavenumbers for the $m=1$ mode
is therefore, approximately, (\cite{Acheson:1978}, \cite{Spruit:1999}):

\begin{equation}
\frac{\sigma}{\eta} > n^2 > \frac{N^2}{\omega_{\rm A}^2 r^2},
\label{eq:minandmaxofn}
\end{equation}

where $r$ is some measure of the size of the field in the horizontal direction.
This is the condition in the absence of thermal diffusion ($\kappa=0$). A 
complication arises here because it turns out that in the presence of both thermal 
and magnetic diffusion, the limit $\kappa/\eta \rightarrow 0$ is a singular one 
with respect to the instability condition: there exist double diffusive instabilities 
which are absent in the case $\kappa=0$. Since there is always some thermal 
diffusion due to numerical effects, we should not expect the simulations to 
reproduce (\ref{eq:minandmaxofn}) accurately. This is discussed further below.

For a $p=1$ field with $\mathbf{B}=(0,B_0\varpi/\varpi_0,0)$ 
in the slowly-rotating case $\Omega\ll\omega_{\rm A}$ we have unstable 
wavenumbers $n$, and hence instability, if (using Eq. (\ref{eq:growthratelo}))

\begin{equation}
\omega_{\rm A}^3 > \frac{\eta N^2}{r^2}.
\label{eq:nin}
\end{equation}

Instability is thus suppressed at low field strengths. The instability
condition is a function of the meridional coordinates ($\varpi,z$).

\subsection{The effect of thermal diffusion}

Thermal diffusion has the effect of reducing the stabilising effect of the
density stratification, allowing instability for a larger range of vertical
wavenumbers. This is a general effect, not only for the magnetic instabilities
discussed here. It was first noted in the astrophysical context by Zahn (1974).
It is important in stars since the thermal diffusion (measured by the diffusivity,
with units area per unit time) is much faster than other damping effects (like 
magnetic and viscous diffusion). In this case there exist length scales large
enough that these other damping effects are negligible on the time scale
on which the instability operates, but at the same time small enough that
thermal diffusion can wipe out the temperature fluctuations due to 
vertical displacement against the stable thermal stratification. This led to
Zahn's (1974, 1983) formulation of shear instability in stellar interiors, widely
used in stellar evolution calculations.

The effect can be incorporated by replacing the buoyancy frequency
frequency $N$ by a value $\tilde{N}$ given by
\begin{equation}
\tilde{N}^2 = \frac{N^2}{1+\tau_I/\tau_T},
\label{eq:thermdiff}
\end{equation}
where $\tau_I$ is the timescale of the instability, i.e. $1/\sigma$,
and $\tau_T$ is the thermal timescale, equal to $1/\kappa n^2$. [This is correct 
by order of magnitude for all instabilities, and can be made more exact for any 
given one]. This should manifest itself in a decrease of the minimum unstable 
wavenumber given in (\ref{eq:minandmaxofn}) above; 
it has exactly the same effect as a reduction in the acceleration of gravity $g$. 

\subsection{The model}

To simulate conditions appropriate to a stratified stellar interior,
without simulating the star as a whole, we make use of the fact that
the Tayler instability 
is always present in at least some region near the magnetic
axis. It has a low threshold, being suppressed only at very low field
strengths where the combined effects of vertical stratification and magnetic 
diffusion suppress the instability (cf. Eq. (\ref{eq:nin})). We use a local plane-parallel
approximation of the star, with the direction of gravity parallel to the
magnetic axis. This is a good approximation for a small region near
the axis at the low field strengths of primary interest, since the
(meridional components of the) unstable wavenumbers are then high
(cf. Eqs. (\ref{eq:minandmaxofn}) and (\ref{eq:nin})).

The ideal gas equation of state, measuring temperature in units such
that the molar gas constant divided by the molecular mass is unity, is
\begin{equation}
P = \rho T \qquad\mbox{and}\qquad e=\frac{T}{\gamma-1}.
\label{eq:state}
\end{equation}
The momentum equation:
\begin{equation}
\frac{{\rm D}\mathbf{u}}{{\rm D}t} = -\frac{1}{\rho}\mathbf{\nabla} P +
\frac{1}{\rho}\mathbf{J\times B} + \mathbf{g} + 2\mathbf{u \times \Omega}.
\end{equation}
{\mk Conservation of mass:}
\begin{equation}
\frac{{\rm D}\rho}{{\rm D}t} = -\rho\mathbf{\nabla .u}.
\end{equation}
Specific internal energy: (thermal conduction and Ohmic heating)
\begin{equation}
\frac{{\rm D}e}{{\rm D}t} = -T\mathbf{\nabla .u} + \gamma\kappa\frac{1}{\rho}\mathbf{\nabla .}(\rho \mathbf{\nabla} e) + \eta\frac{1}{\rho} J^2.
\label{eq:De/Dt}
\end{equation}
Electric current:
\begin{equation}
\mathbf{J} = \mathbf{\nabla \times B},
\end{equation}
and the induction equation:
\begin{equation}
\frac{\partial\mathbf{B}}{\partial t} = -\mathbf{\nabla \times} (\eta \mathbf{J} - \mathbf{u\times B}),
\label{eq:indu}
\end{equation}

where in the above equations, the velocity field is
denoted by $\mathbf{u}$, the differential operator $D/Dt \equiv
\partial /\partial t + \mathbf{u.\nabla}$, $\eta$ and $\kappa$ are the
magnetic and thermal diffusivities respectively, $e$ is specific
internal energy and $\mathbf{g}$ is
gravitational acceleration. We ignore fluid viscosity.

In the calculations, we have left out the ohmic heating term $\eta
J^2/\rho$ from the energy equation (\ref{eq:De/Dt}). In a real star, there
are other sources and sinks of energy (radiation) which together
determine its thermal equilibrium structure. Since 
the magnetic energy $B^2$ is small, thermal changes due to Ohmic 
dissipation are slow compared with the (Alfv\'enic) time scale of
interest here, 
it is more consistent to leave out the gradual heating by Ohmic diffusion
together with these other sources of slow thermal change.

\section{3D MHD simulations}

\subsection{The numerical code}

We use a three-dimensional MHD code developed by Nordlund \&
Galsgaard (1995). The code uses a Cartesian coordinate system.
This has the advantage, over a code using cylindrical
coordinates, not only that the code is significantly simpler, but that
it avoids the coordinate singularity on the axis, known to cause
serious problems in essentially all grid-based methods in cylindrical
or spherical coordinates. The disadvantage is that an intrinsically
round peg (the toroidal field) has to be fitted into a square
hole. The penalties are some waste of computing time (the grid points
in the corners being unused), and a small amount of startup noise
because a good initial equilibrium state is harder to construct.

The code uses a staggered mesh, so that variables are defined at
different points in the gridbox. For example, $\rho$ is defined in the
centre of each box, but $u_x$ in the centre of the face perpendicular
to the x-axis, so that the value of $x$ is lower by
$\frac{1}{2}dx$. Interpolations and spatial derivatives are calculated
to fifth and sixth order respectively. The third order
predictor-corrector time-stepping procedure of Hyman
(1979) is used.

{\mk For numerical stability, the code contains high-order damping terms. These
were however
switched off in this study} -- we were interested in the early (and
therefore linear) stage of instability growth, before displacements and
velocities become large enough to necessitate any stabilisation.

\subsection{Definition of the stratification}
\label{sec:strat}

We want to investigate, amongst other things, which vertical 
wavenumbers are unstable. This is achieved most easily if the growth 
rates of the instabilities, and indeed all timescales,
are independent of $z$. To this end $T$ has to be made independent of $z$,
so that the sound crossing time over the width of the box is independent of 
height. 
The stratification is thus approximated by that of an isothermal
atmosphere. This is sufficient for the present purpose, since it
allows the inclusion of the essence of the stabilising effect of buoyancy.

The equilibrium magnetic field $\mathbf{B}$ is then made
dependent on $z$ in such a way that the ratio of thermal and magnetic
energy densities, $\beta=e\rho\, 8\pi/B^2$, and therefore
the Alfv\'{e}n speed, is independent of $z$.

\subsection{Initial magnetic configuration and equilibrium}

Let $(\varpi,\phi)$ denote polar coordinates on a horizontal  plane. 
We consider a purely azimuthal field of the form
\begin{equation}
\mathbf{B} = B_0 \left(\frac{\varpi}{\varpi_0}\right)^p
e^{-\varpi^2/\varpi_0^2} e^{-z/2H} \mathbf{e_\phi}\label{eq:binit},
\end{equation}
where $\varpi_0$ is a length constant, set to a quarter of the
horizontal size of the computational box, and $B_0$ is a constant 
which determines the strength of the magnetic field. The exponential 
containing z keeps the value of $\beta$ independent of $z$ (see Sect. 
\ref{sec:height}). The instability is in the centre, where $\varpi/\varpi_0 
\ll 1$. The exponential containing $\varpi$ is present 
to make the axisymmetric field fit inside a square computational box. As the 
field evolves during the instability, it will spread outwards so that some 
margin has to be left empty around the region containing the field. The 
results show that this has been successfully avoided, at least during the 
linear phase of growth.

To ensure that the initial state is close enough to equilibrium to
convincingly measure the growth of the instability, the pressure, density and 
temperature fields have to be adjusted to the magnetic force brought about 
the field given in (\ref{eq:binit}).

Let
\begin{equation}
\rho(\varpi,z)=\rho_0 W(\varpi) e^{-gz/T_0}
\quad\mbox{and}\quad T(\varpi)=T_0 D(\varpi).
\end{equation}
The equilibrium conditions (Eqs. (\ref{eq:state})--(\ref{eq:indu}) with $\partial_t= 
{\mathbf v}=\eta=0$) are then satisfied for the axisymmetric field
given by (\ref{eq:binit}) provided that
\begin{equation}
W(\varpi)=1 + \frac{B_0^2}{4\rho_0 T_0} e^{-2\varpi^2/\varpi_0^2}
        \qquad\mbox{and}
\end{equation}
\begin{equation}
D(\varpi)=1 - \frac{2\varpi^2/\varpi_0^2}{1+\frac{4\rho_0 T_0}{B_0^2}e^{2\varpi^2/\varpi_0^2}}
\end{equation}
for the case $p=1$,
or, if $p=2$,
\begin{equation}
W(\varpi)=1 + \frac{B_0^2}{8\rho_0 T_0} \left(1+2\frac{\varpi^2}{\varpi_0^2}\right) e^{-2\varpi^2/\varpi_0^2}
        \qquad\mbox{and}
\end{equation}
\begin{equation}
D(\varpi)=1 - \frac{4\varpi^4/\varpi_0^4}{1+2\frac{\varpi^2}{\varpi_0^2}+\frac{8\rho_0 T_0}{B_0^2}e^{2\varpi^2/\varpi_0^2}}.
\end{equation}

{\mk Magnetic diffusion will} destroy 
the equilibrium even in the absence of dynamic instability. This should not 
be problematic as long as the timescale over which the instability becomes
visible is shorter than the timescale over which the equilibrium is
lost in this way. This condition is comfortably fulfilled in the
simulations presented here, because the instability operates on a small length
scale. Magnetic diffusivity of a value which is relevant for the instability 
has little effect on the much larger length scale of this initial field 
configuration.

\subsection{Boundary conditions}

Under the conditions prevailing in stellar interiors, the expected length
scale of the instabilities is very small (at least in the radial direction). As in the 
case of hydrodynamic instabilities, therefore, different parts of the star are 
in effect disconnected from each other as far as the instability is concerned. The 
computational box can then be taken to cover only a small part of the star, 
as is also done in studies of magnetic instability in thin accretion discs (e.g.\ 
Hawley et al. 1995).
This raises a problem with the boundary conditions, since the box boundaries 
are special locations in the computational volume that have no counterpart 
in the real star. With rigid boundaries, spurious phenomena such as boundary 
layers would affect the results. Such effects are minimised by using periodic 
boundary conditions: there are then no special locations in the computational 
box. 

\subsubsection{Horizontal}
Periodic conditions in the $x$ and $y$ directions are implemented by copying 
unknown values of $P,T,\rho$ and $\mathbf v$ outside the box boundaries 
from their values at $x\pm L$, $y\pm L$ where $L$ is the width of the box.
For the magnetic field, 
the sign of $\mathbf{B}$ is reversed in the copying process, so that there are 
no current sheets at the boundaries. In any case, with the
field used, the values at the boundaries are very small
compared to the values at the centre of the box.

\subsubsection{Vertical}
\label{sec:height}
In the vertical direction periodic conditions are
also possible by making use of an invariance of the equations for the
case of an isothermal stratification. By this symmetry, a shift in
height is equivalent to multiplication of the variables by constant
factors. In this way the variables can be scaled from the bottom to
the top of the computational box. Thus, when interpolating or 
differentiating across the top and bottom of the computational box, 
we multiply or divide the density by a factor $exp(L_z/H)$ (where $L_z$ is 
the height of the box and $H$ is the scale height). These `scaled periodic'  
boundary conditions are the appropriate equivalent of periodic  conditions 
for a stratified medium. 

Unfortunately, it turns out that this procedure does not
work in its simplest form. Small perturbations launched by the initial
disequilibrium propagate vertically, as a mixture of sound waves and
internal gravity waves. The upward propagating components of a sound
wave will grow (in the limit where the wavelength is
much smaller than the scale height) at a rate of the order of the
buoyancy frequency (differing from it by a factor $1.02$ if
$\gamma=5/3$). If the wave leaving the top of the box is fed
back in at the bottom, this amplification goes on indefinitely.

Though the initial perturbations can be made small, the
exponential growth of the upward-travelling waves causes a problem if
this growth is faster than that of the magnetic instability
under study. Since we want to operate in the
regime where $N \gg \omega_{\rm A}$, this will indeed be the
case. Engineering fixes like artificial damping terms designed to affect 
sound waves would not greatly help, since internal gravity waves will 
also grow in this way at a comparable rate.

Many methods were tried to extinguish these waves. The most
satisfactory solution found was to divide the volume vertically
into two halves, with gravity pointing downwards and upwards in 
the top and bottom halves of the box respectively. A sound wave is 
still amplified within one half, but will not see any net increase in
amplitude over the course of a journey through the whole box.
The physics of the magnetic instability is replicated in the two
halves, so that the penalty is a factor of two in computational cost.

\subsection{Initial perturbations}
All of the runs were started with a small perturbation to the velocity
field. This could either be given at all values of $m$ and $n$, or
just to one particular mode. There was some uncertainty as to the
precise form of the perturbation required; how it should vary as a
function of $\varpi$, whether it should include a vertical component
and in what sense the phases of the two horizontal components should
vary with $z$ were all unanswered questions. Several different
formulations were tried, producing, reassuringly, virtually identical
results. For the runs appearing here, the initial perturbation had its maximum
at $\varpi=0$ and died away at increasing $\varpi$ so that at
$\varpi=\varpi_0$ it was almost zero, the vertical component of the
velocity perturbation was zero and for the $m\neq 0$ modes, the phase
in the $\phi$ direction went from $0$ to $2\pi$ over the course of one
vertical wavelength so that the cross-section remained the same at all
$z$ but was rotated about the axis once per wavelength. The initial
perturbation is therefore of the form:

\begin{eqnarray}
\mathbf{v} & = & \sum_{m=0,n} V_0^n
\frac{\varpi}{\varpi_0}\exp\left(-\frac{5}{2}\frac{\varpi^2}{\varpi_0^2}\right)
\cos(nz)\mathbf{e}_{\varpi} \nonumber\\
& + & \sum_{m\neq0,n} V_m^n
\exp\left(-3\frac{\varpi}{\varpi_0}-\frac{\varpi^2}{\varpi_0^2}\right)
[\cos(nz-m\phi)\mathbf{e}_{\varpi} \nonumber\\
&   & \qquad\qquad\qquad\qquad\qquad\quad\quad  +\sin(nz-m\phi)\mathbf{e}_{\phi}].
\end{eqnarray}

{\mk The amplitude of the initial velocity perturbation $V_m^n$ was
given a value $3\times10^{-5}$, which corresponds to a fraction
$2\times10^{-5}$ of the sound speed or $2\times10^{-4}$ of the
Alfv\'en speed. This value is large enough that the numerical
perturbation is small in comparison, while being small enough to
follow the linear phase of instability growth for several growth time-scales.}

\subsection{Accessible parameter values}
The separation of time scales $\omega_{\rm A} \ll N$ and $\Omega \ll N$ appropriate
for stars is also convenient for the analytical stability
analysis. Such widely differing time scales are problematic for
numerical simulations, however, since the time step will be set
by the shortest time scale, the sound crossing time (which is of the
order of $1/N$), while the phenomena of interest happen on the slower
magnetic time scale. In the first set of calculations reported below
we have set $\Omega = 0$, and the ratio $\omega_{\rm A}/N$ is of order
$0.01-0.1$. For such values, the predicted growth rates are already
close to their asymptotic ($\omega_{\rm A} \ll N$) value, so not much would
be gained from more expensive calculations with lower field
strengths. The effect of rotation is investigated in Sect. \ref{sec:rot}.

\section{Results}

To test the analytic predictions and the behaviour of the code a series of 
test cases was studied, starting with a simple unstratified medium with 
adiabatic conditions. Physical ingredients are then added concluding with 
the case of most interest for a stellar interior, in which stratification, 
magnetic and thermal diffusion are included.

\subsection{Dependence of stability on $p$}
\label{sec:pandm}

Two initial field configurations, $p=1$ and $p=2$, were tried. A
resolution of $48\times48\times48$ was used for these runs, and
the computational box is a cube of side $2\pi$. A zero value was used for
$g$, and $B_0=1.28$. As in all runs $\gamma=5/3$ and
$\varpi_0=\pi/2$. This produces a maximum value of the magnetic energy density of
one tenth of the thermal energy density, (i.e. $\beta=10$). This is 
still in the weak-field approximation ($v_{\rm A}\ll c_{\rm s}$).

\begin{figure*}
\includegraphics[width=0.33\hsize]{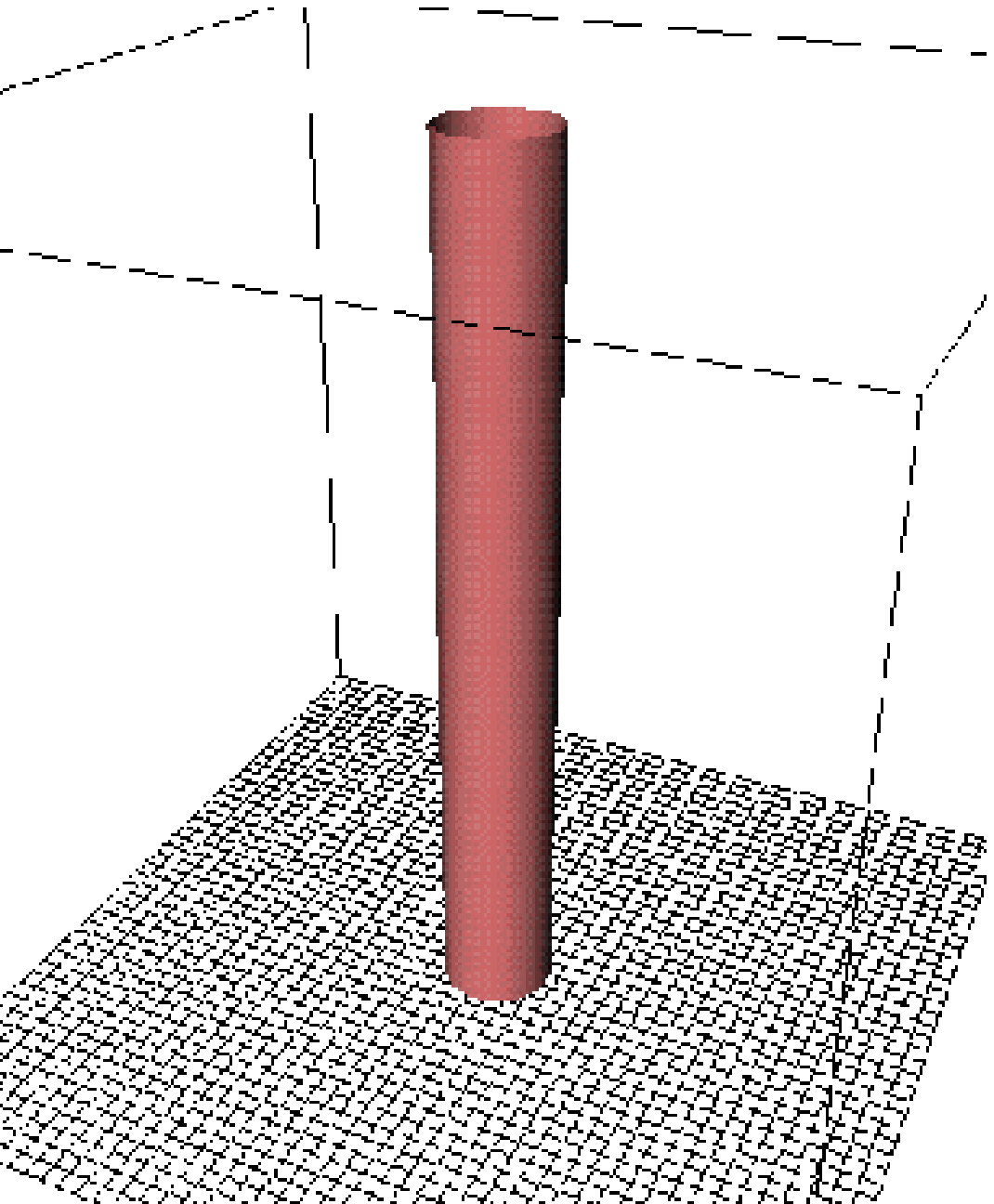}
\includegraphics[width=0.33\hsize]{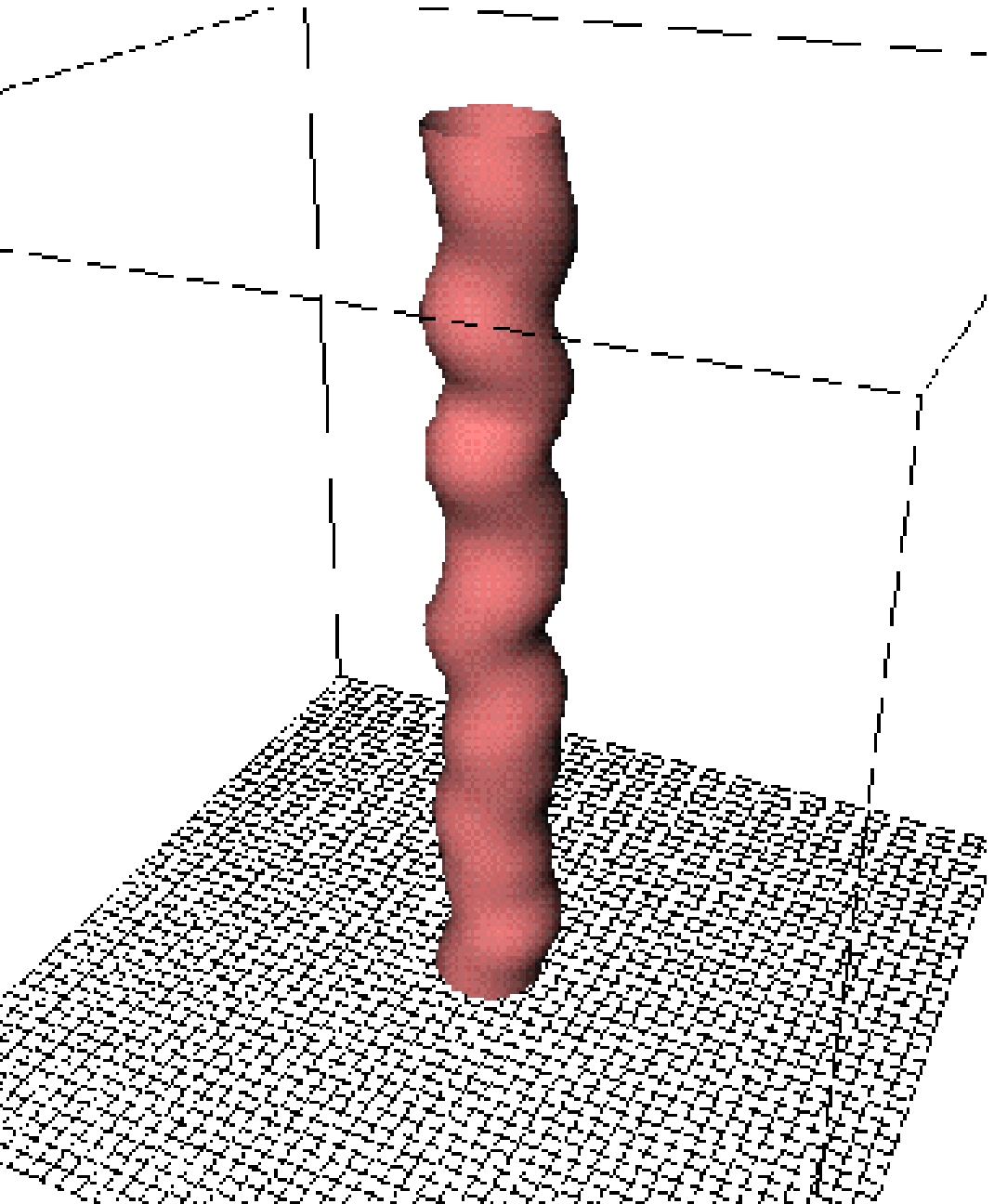}
\includegraphics[width=0.33\hsize]{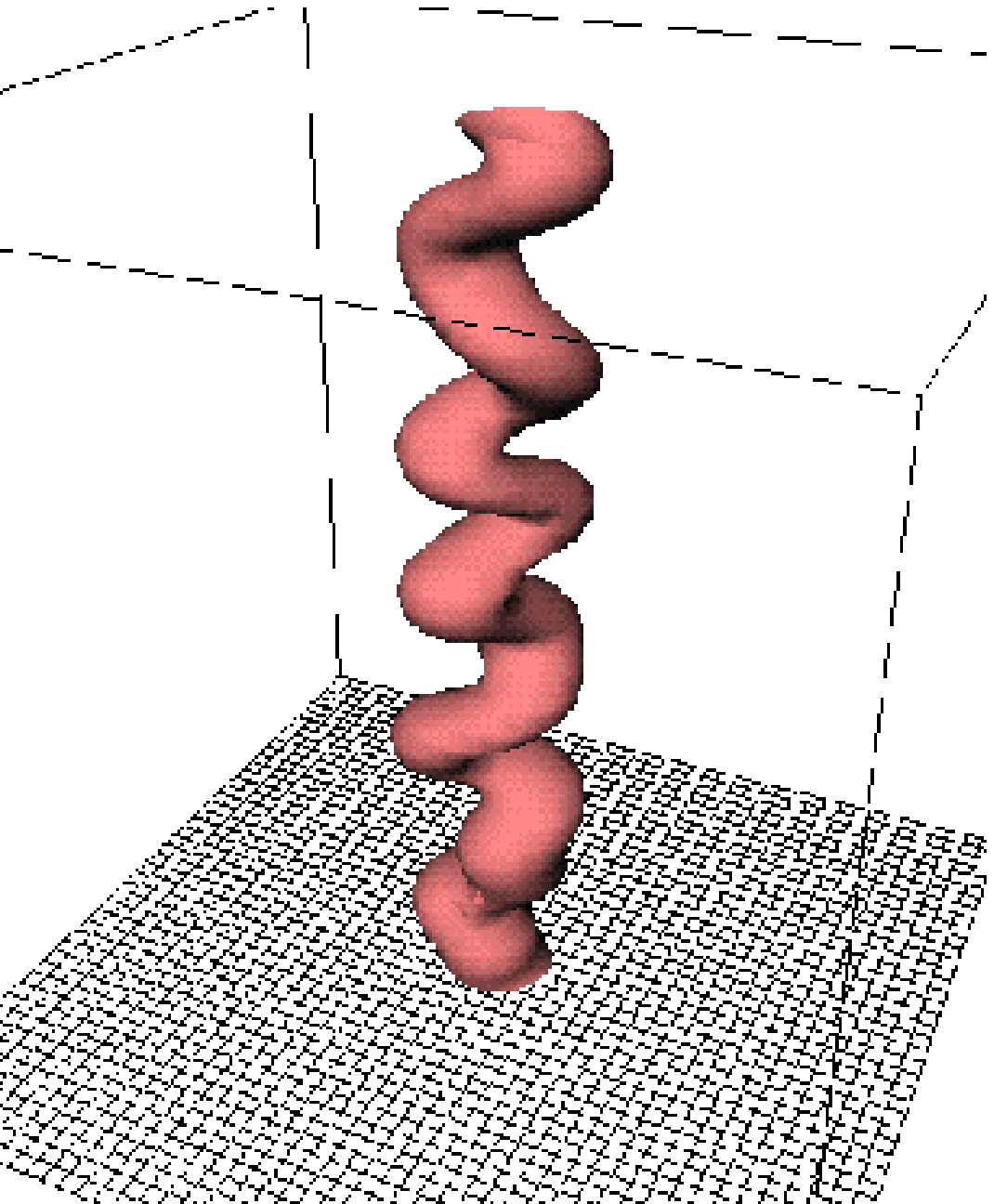}
\\
\includegraphics[width=0.33\hsize]{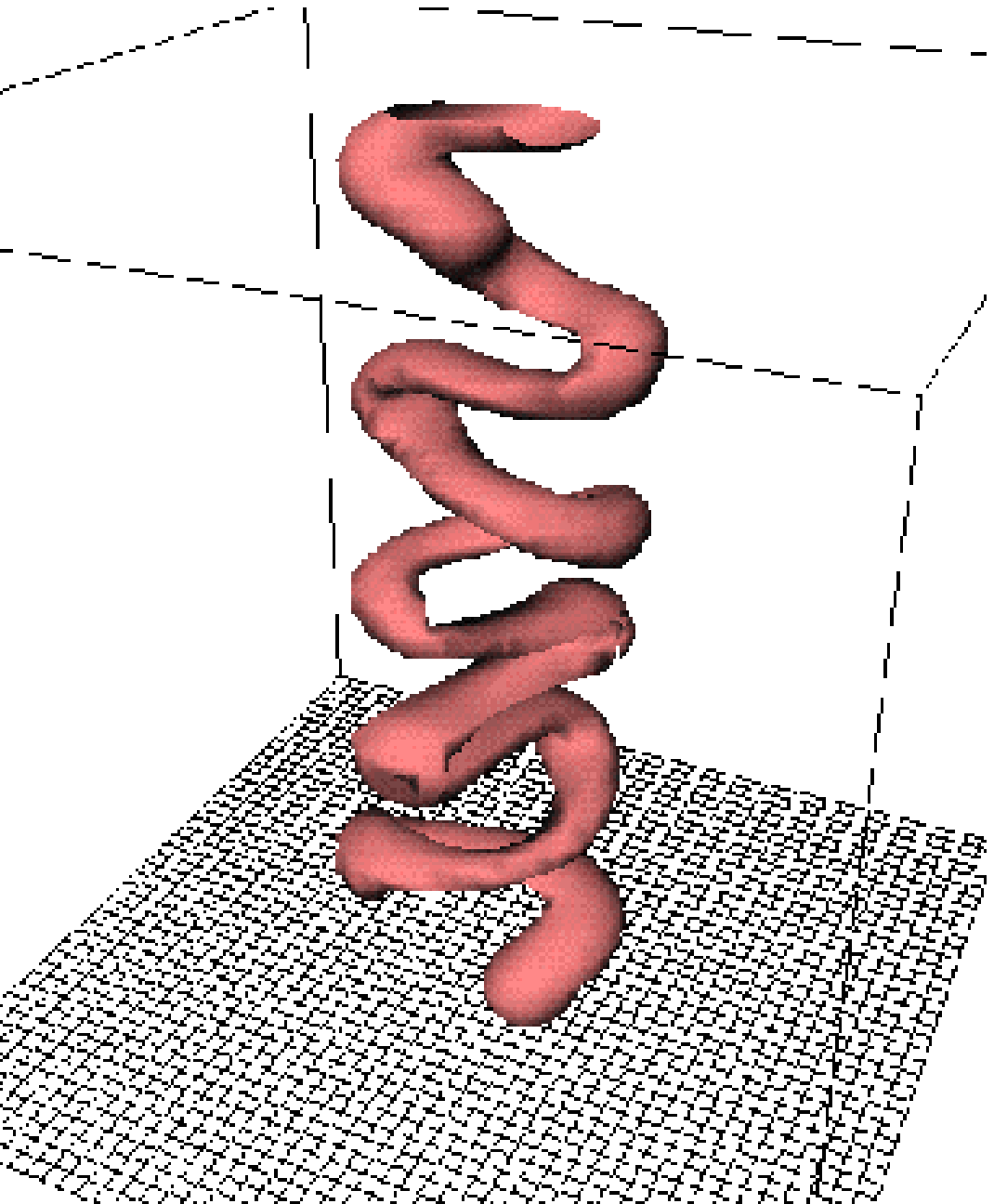}
\includegraphics[width=0.33\hsize]{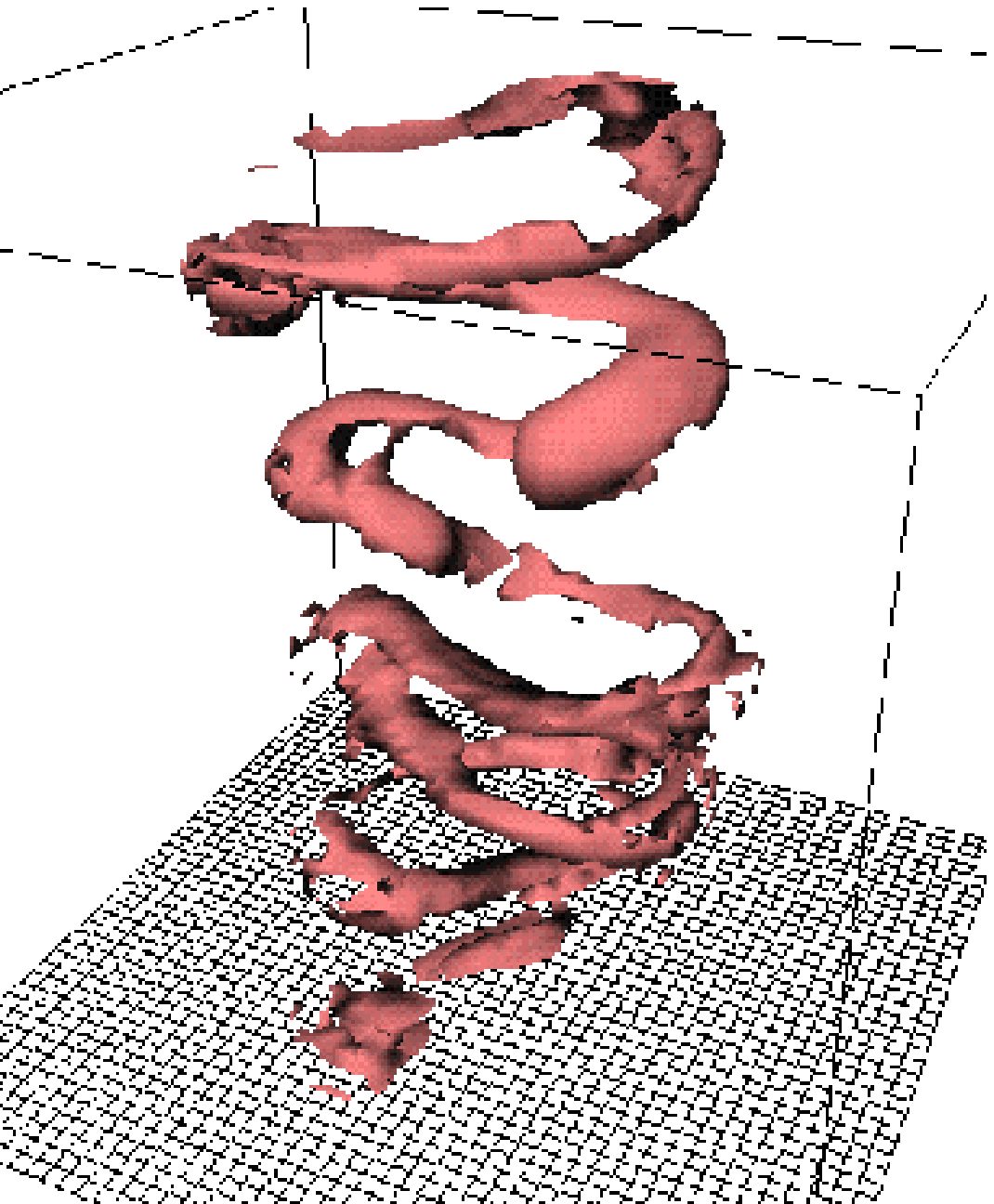}
\includegraphics[width=0.33\hsize]{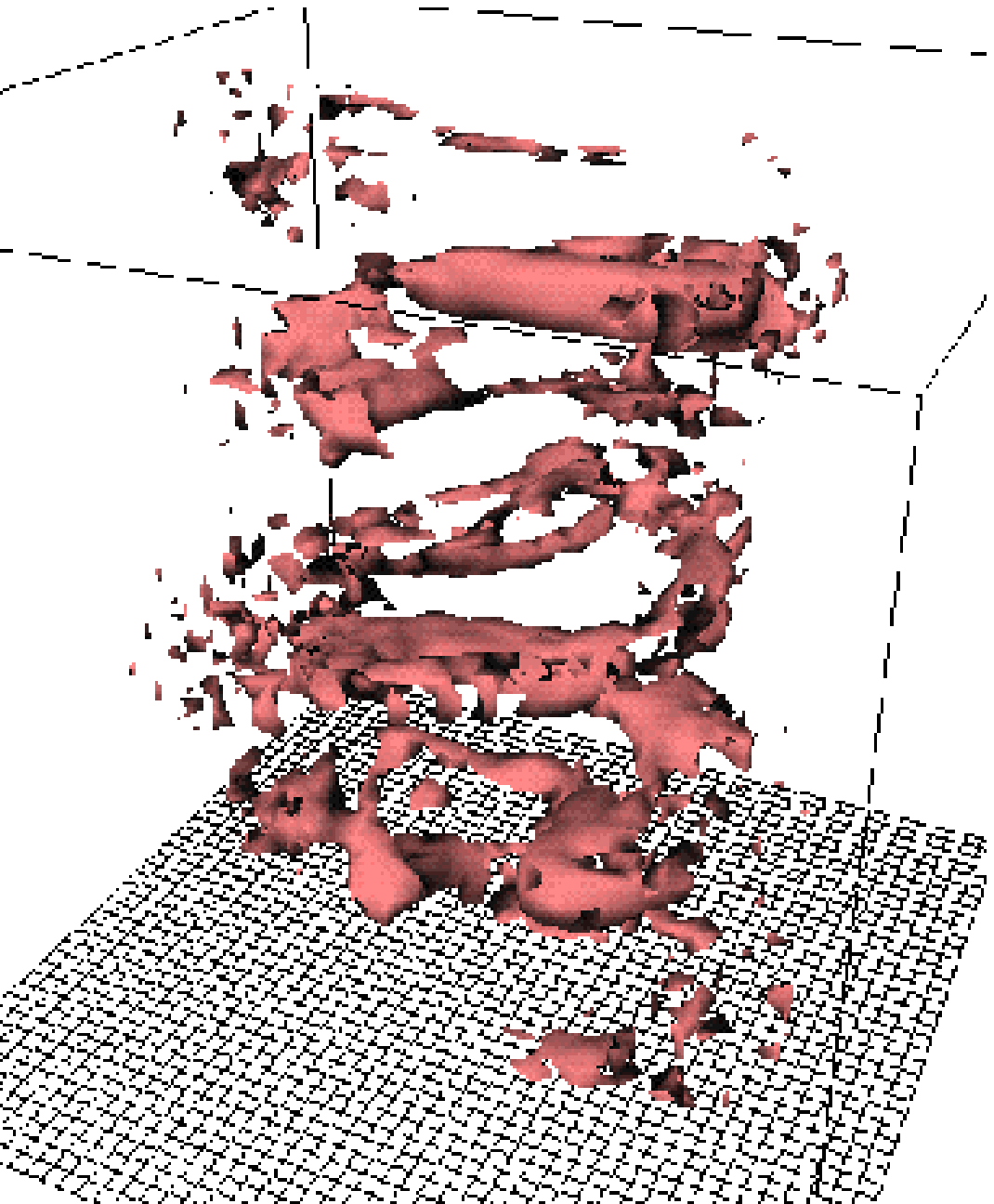}
\caption{Following the evolution of instabilities -- surfaces of constant initial $\varpi=\pi/10=\varpi_0/5$ at
roughly regular time intervals:
$t=6.30,9.07,10.86,12.52,14.56,16.20$.}
\label{fig:animation}
\end{figure*}

To follow the development of the instability in time, it was found useful
to calculate the Lagrangian displacement field. This was computed by 
adding a vectorial tracer field $\chi\,=\,(\chi_x$, $\chi_y$, $\chi_z)$ to the 
code. The initial values of this field were simply the coordinates $x_0$, $y_0$
and $z_0$ at $t=0$, and it was evolved according to the equation
${\rm D}\chi/{\rm D}t=0$, making use of the same advection routines as in the rest
of the code. The displacement field ($\mathbf{\xi}$) can be found by 
subtracting the tracer field from its initial value. This Lagrangian displacement 
field was used for the representations shown in Fig. \ref{fig:animation}.

In Fig. \ref{fig:animation} we can see the evolution of the
displacement field near the magnetic axis: it shows how a fluid volume
that was initially a cylinder around the axis evolves in time under
the influence of the instability of the magnetic field, in the $p=1$
case. The surfaces shown are surfaces of constant $\chi_x^2+\chi_y^2$. 
Some magnetic diffusion ($\eta=3\times10^{-3}$) has been added
to remove the higher spatial frequencies and give the picture a more
pleasing look. Is it clear that we are looking at an $m=1$
instability since the plasma is displaced in the horizontal direction
without losing its circular cross-section.

\begin{figure*}
\includegraphics[width=0.5\hsize]{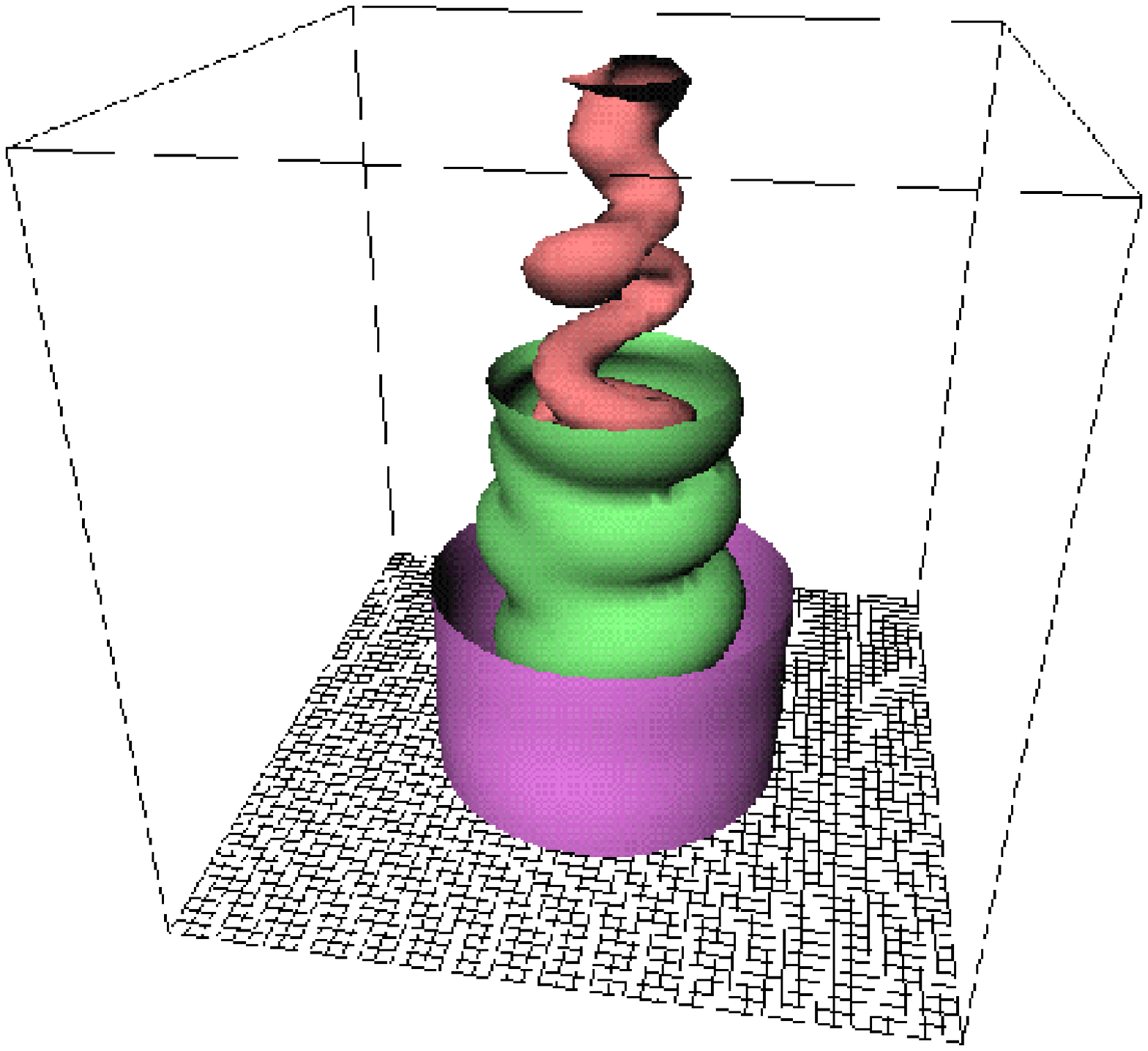}
\includegraphics[width=0.5\hsize]{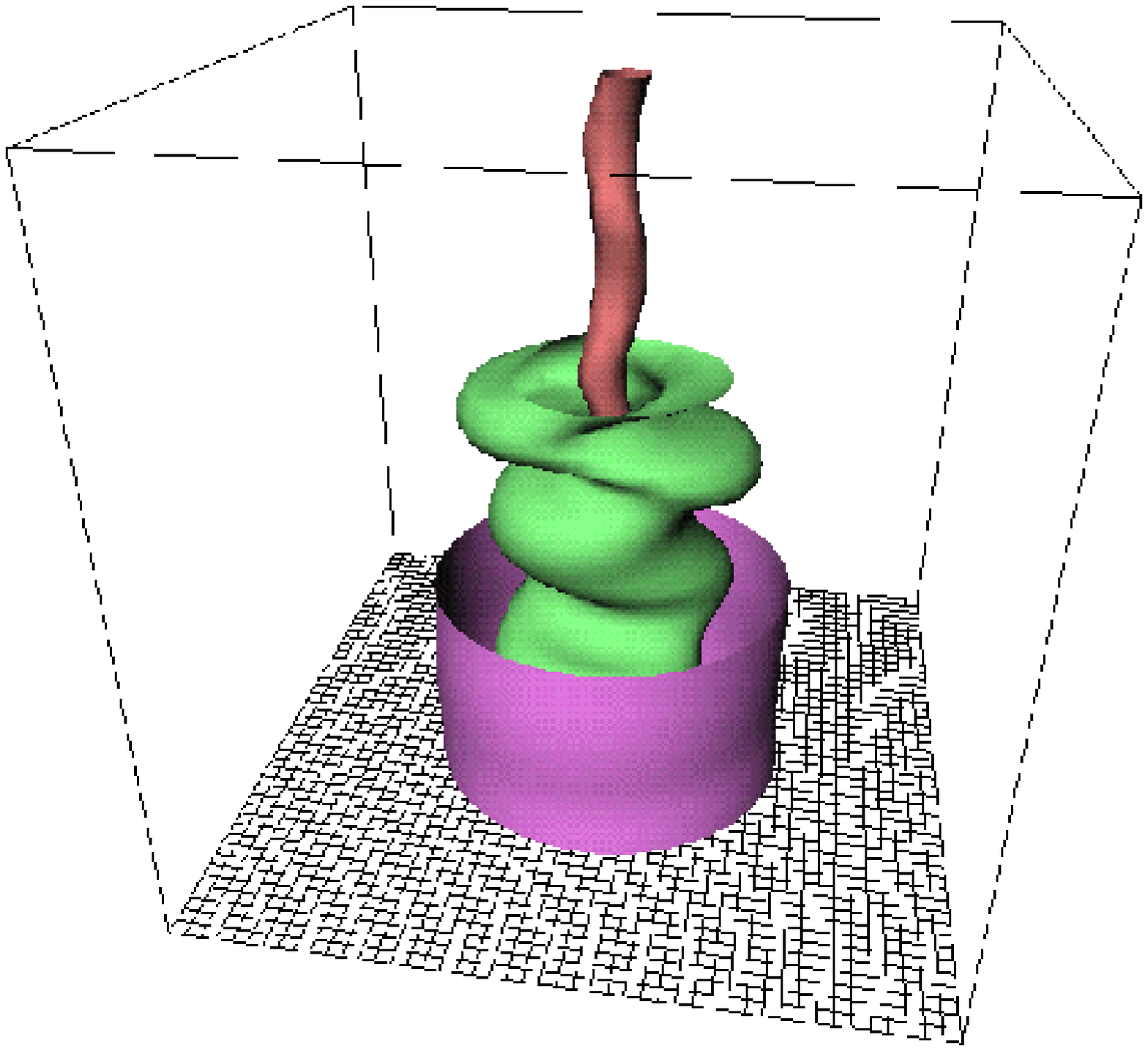}
\caption{
Two runs, identical except for the value of $p$, which is equal to $1$
and $2$ in the left and right hand figures respectively. The left hand figure is
the same as the fourth frame of the sequence in Fig.
\ref{fig:animation}, i.e. at $t=12.52$. In the right hand figure,
$t=15.29$. Plotted are three surfaces of constant initial $\varpi$
(i.e. constant $\chi_x^2+\chi_y^2$) inside one another. It is clearly visible that the
instability is on the axis in the case of $p=1$, but away from the
axis when $p=2$.}
\label{fig:springonion}
\end{figure*}

In Fig. \ref{fig:springonion} we see the $p=1$ and $p=2$ runs compared;
three surfaces of constant $\chi_x^2+\chi_y^2$ have been plotted, two of which have been
partly cut away. These confirm that in the $p=1$ case, the instability grows most
quickly on the axis, and that in the $p=2$ case, it grows most quickly
away from the axis. This should be no surprise as here the local
Alfv\'{e}n frequency $\omega_\mathrm{A}$ (see Eq. (\ref{eq:defofoa})) is greatest.

\begin{figure}
\includegraphics[width=1.0\hsize,angle=0]{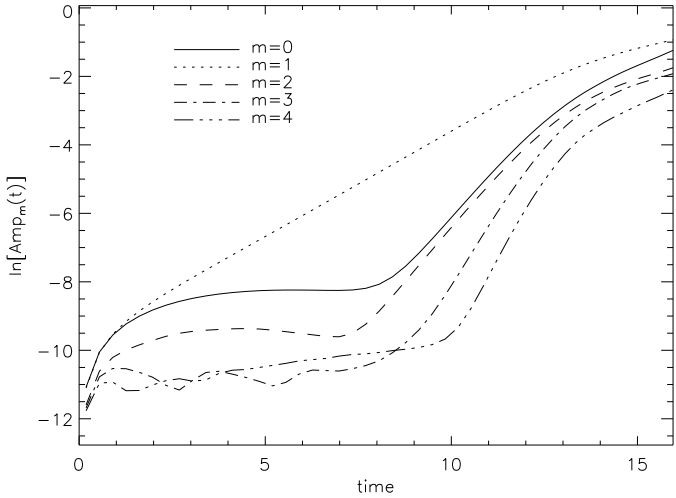}
\caption{Displacement amplitudes of different $m$ modes, when
$p=1$. Adiabatic ($\kappa=\eta=0$) unstratified case.}
\label{fig:p1amp}
\end{figure}

It is possible to perform a Fourier decomposition of the
displacements in the $\phi$ dimension, enabling us to see the amplitudes of the
different $m$ modes separately. We can write the displacement
$\mathbf{\xi}$ as

\begin{equation}
\mathbf{\xi}(\varpi,\phi,z,t) = \frac{1}{2}\mathbf{A}_0 (\varpi,z,t)
+ \sum_{m=1}^{\infty} \Re (\mathbf{A}_m (\varpi,z,t) e^{im\phi})
\end{equation}
and calculating
\begin{equation}
{\rm Amp}_m(t)={\left( \frac
  {\int_{\varpi=0}^{\varpi_\mathrm{max}} \int_{z=0}^{L_z} \mathbf{A}_m^\ast. 
    \mathbf{A}_m 2\pi \varpi \mathrm{d}\varpi \mathrm{d}z}
  {\pi \varpi_\mathrm{max}^2 L_z}
\right) }^{\frac{1}{2}}
\label{eq:intovervarpi}
\end{equation}

should give us a suitable measure of the amplitudes of each $m$
mode. A value of $\varpi_{\rm max}=0.5 \varpi_0$ is used, since this is
the region of interest, where the field strength is roughly
proportional to $\varpi^p$. In any case, choosing a different value
doesn't have any significant affect on the results. This quantity
${\rm Amp}_m$ (or rather, its logarithm) is plotted in Fig. \ref{fig:p1amp} 
for the aforementioned $p=1$ run. We can see that the $m=1$ mode is the 
only unstable one when $p=1$.

\begin{figure}
\includegraphics[width=1.0\hsize,angle=0]{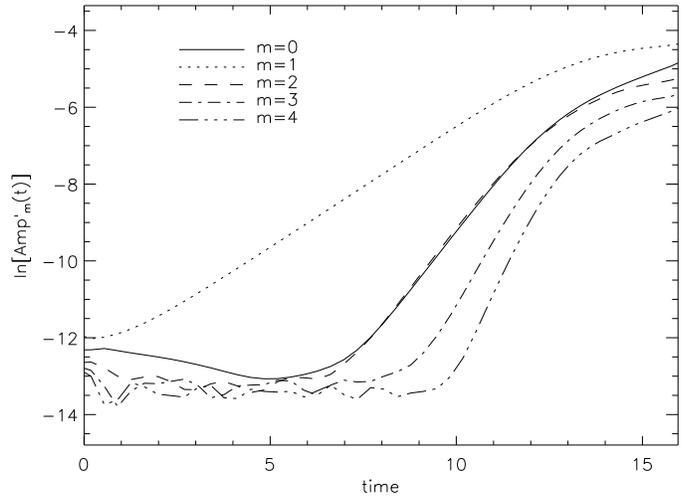}
\caption{Velocity amplitudes of different $m$ modes, when $p=1$. Adiabatic unstratified case.}
\label{fig:p1velamp}
\end{figure}

It is possible to do this Fourier decomposition on not just the
displacement field but also on the velocity field, giving an analogous
amplitude ${\rm Amp}^\prime_m$. This quantity is plotted in Fig.
\ref{fig:p1velamp}. Whereas the amplitudes in the displacement field
begin at zero, the amplitudes in the velocity field begin with a
finite value, since we are giving the plasma an initial perturbation to the
velocity field.

\begin{figure}
\includegraphics[width=1.0\hsize,angle=0]{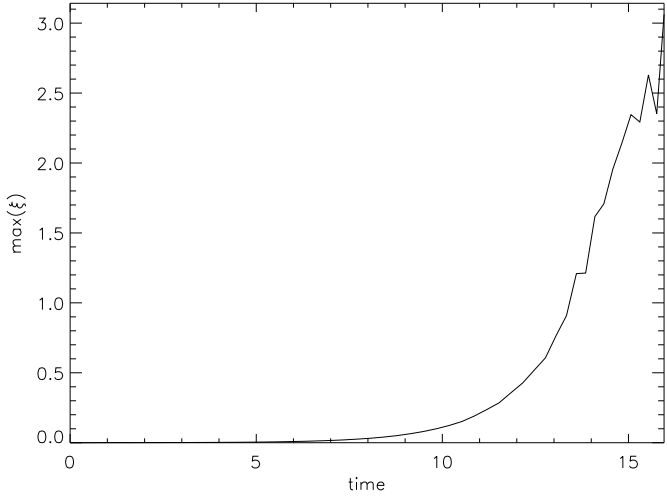}
\caption{Maximum displacement, when $p=1$. The computational box is a cube of side
$2\pi$, and the shortest wavelengths of the instability may be just a
twenty-fourth of this distance. Adiabatic unstratified case.}
\label{fig:p1maxdisp}
\end{figure}

At some point the other modes begin to grow also, presumably when the
$m=1$ mode enters the non-linear phase. To test this hypothesis, we can
consider that the smallest vertical wavelength present is two 
grid boxes, equal to a distance 0.26, so the minimum
value of $1/n$ will be $\lambda_\mathrm{min}/2\pi = 0.042$. The non-linear phase begins once the
displacements become comparable to this. Fig. \ref{fig:p1maxdisp}
shows the maximum value in the displacement field, i.e. the maximum
displacement from equilibrium reached anywhere in the computational box -- 
we can confirm that this value reaches $\lambda_\mathrm{min}/2\pi$ at roughly the same time as the
$m\not=1$ modes begin to grow.

\begin{figure}
\includegraphics[width=1.0\hsize,angle=0]{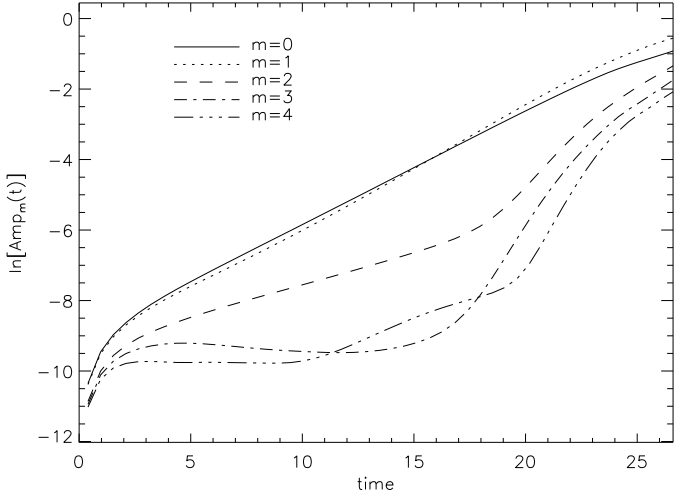}
\caption{Displacement-field amplitudes of different $m$ modes, when $p=2$. Adiabatic unstratified case.}
\label{fig:p2amp}
\end{figure}

In the case of $p=2$, the instability grows more quickly away from the
axis. We have already seen this in Fig. \ref{fig:springonion}. This
complicates the analysis, as the instability is strongest in the
region where the initial field deviates from proportionality to
$\varpi^p$. This is to be expected, since the growth rate is of order 
$v_{\rm A}/\varpi$ which is proportional to $\varpi$ if $p=2$. However, it 
should be possible to use data from the inner region only, before it is
contaminated by non-linear development from the outer region.
Fig. \ref{fig:p2amp} is the equivalent of Fig.
\ref{fig:p1amp} for the $p=2$ case. We can see that the $m=0$ and $m=2$ modes
are growing, as well as the $m=1$ as in the previous case.

All subsequent discussion is limited to the physically more likely $p=1$ case.

\subsection{Stratified and non-stratified runs and vertical wavenumber}
\label{sec:stratandnon}

For the runs described above, we were after just a qualitative result,
as opposed to the quantitative result required from the following
runs, which had consequently to be set up in a slightly different
way.

In some cases we were interested in the behaviour at different
vertical wavelengths. We therefore performed a second Fourier decomposition of the
displacement (or velocity) field, this time in the vertical
direction. This yields the amplitude at each value of $m$ and $n$
(values of the latter correspond to wavelengths from the height of the
box down to two grid spacings):
\begin{equation}
\mathbf{A}_m(\varpi,z,t)=\frac{1}{2}\mathbf{C}_{m0}(\varpi,t) +
\sum_n \Re (\mathbf{C}_{mn}(\varpi,t) e^{inz}).
\end{equation}
As before (see Eq. (\ref{eq:intovervarpi})), an integration of the
coefficients $\mathbf{C}_{mn}$ over $\varpi$ can
be done out to a value $\varpi_{\rm max}$, and it is this amplitude
$\mathrm{Amp}_{mn}(t)$ which is used for the following analysis.

In finding the dependence of instability growth at different vertical
wavelengths as a function of some parameter, (e.g. magnetic
diffusion), one has two options: to measure the growth at several
wavelengths in a single run with some certain value of the parameter
in question, or to measure the growth at a single wavelength in
several runs, each with a different value of the parameter. Two
computational restrictions made the second method the most practical:
firstly, it was discovered that at wavelengths comparable or
greater than $\varpi_0$ grew more slowly than expected -- an effect
purely of the dimensions of the computation box, this meant
that the maximum wavelength which could be examined was $\pi/10$, i.e.
one fifth of $\varpi_0$; secondly, the code could not resolve very
short wavelengths - the minimum wavelength we could look at reliably
was eight grid spacings.

In the cases where stratification was not necessary, therefore, a
vertical resolution of 8 was sufficient, with a box of dimensions
$2\pi\times\pi/10$; a horizontal resolution of 72 proved more than
enough. In cases where it was necessary to hold to the $N \gg
\omega_{\rm A}$ condition, we still looked at a wavelength of $\pi/10$
(corresponding to a wavenumber of $20$), but implemented the scheme described in Sect.
\ref{sec:height}, so that the gravity was in opposite directions in
the two halves of the box. A vertical resolution of 64 was then used
in a box of size $2\pi\times4\pi/5$. All of the following runs were
done with one of these two setups. Since stratification has a
wavelength-dependent effect, i.e. the stratification has a greater
effect on longer wavelengths (see Eq.~\ref{eq:minandmaxofn}),
phenomena which either do not depend on wavelength or only effect the
short wavelengths can be investigated using the non-stratified model
with zero gravity. The first three of the following sections fall into
this category: field strength, rotation and magnetic diffusion. As a
check, the runs with rotation were also done with stratification. The
stratified setup is however required to look, in the last two sections, at the effect of gravity
(obviously) and of thermal diffusion.

\subsection{Dependence of growth rate on field strength}
\label{sec:fs}

The growth rate is expected to be proportional to the field
strength. To see whether this is the case, five runs were done with
$B_0$ equal to $1.28$, $0.404$, $0.128$, $0.0404$ and $0.0128$
respectively. These correspond to minimum values (that is, where the
field is strongest) of 
$\beta$ of $10$ to $10^5$. Magnetic and thermal diffusivity were set
to zero, as was gravity $g$ -- as described in Sect.
\ref{sec:stratandnon} for this non-stratified case, a resolution of
$72\times72\times8$ was used.

\begin{figure}
\includegraphics[width=1.0\hsize,angle=0]{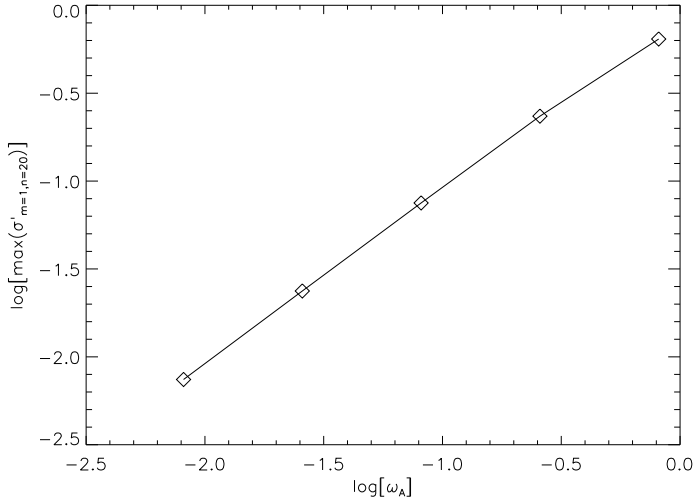}
\caption{The growth rate of the $m=1$ mode is roughly equal to
$\omega_{\rm A}$, which is proportional to field strength. Adiabatic unstratified case.}
\label{fig:bdep}
\end{figure}

The expected growth rate is given by (from Eqs. (\ref{eq:defofoa}) and
(\ref{eq:growthratelo})):
\begin{equation}
\sigma \sim \omega_{\rm A} = \frac{B_0}{\varpi_0 \sqrt{4\pi \rho}}.
\end{equation}
Fig.~\ref{fig:bdep}
shows the maximum growth rate reached in the velocity field of this
$m=1$, $n=20$ mode in the five runs, against the value of $\omega_{\rm A}$.
The velocity field was more convenient
for this purpose than the displacement field, as the growth rate in
the latter begins at infinity (because the perturbation is to the
velocity field) and falls to a steady value, before
falling again once the non-linear stage is reached; the growth rate in
the velocity field begins at zero and rises to a steady value before
falling again -- its maximum value is therefore reached during the
period of steady growth. It can be seen from Fig.~\ref{fig:bdep} that
the growth rate $\sigma$ is proportional to the Alfv\'{e}n frequency
$\omega_{\rm A}$, and is in fact almost equal to it.

The growth rates have been predicted using the approximation that the
magnetic energy density is very much less than the thermal, and that
the Alfv\'{e}n speed is very much less than the sound speed. In the
run with the highest field strength these ratios are only $10$ and
$2.4$, which may explain the slightly lower-than-expected
growth rate here.

\begin{figure}
\includegraphics[width=1.0\hsize,angle=0]{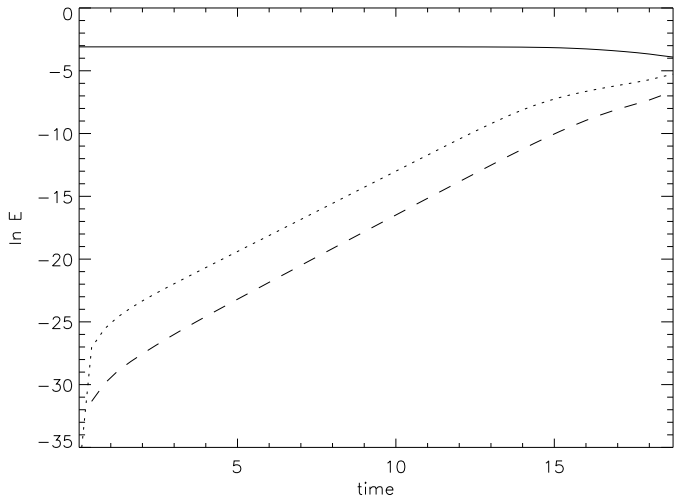}
\caption{{\mk The energy in the three components of the magnetic field. The
solid line represents $B_\phi$, dotted $B_\varpi$ and dashed $B_{\rm
z}$. The two poloidal components are growing exponentially on a
time-scale comparable to the Alfv\'en crossing time. Adiabatic unstratified case.}}
\label{fig:b_comps}
\end{figure}

\begin{figure}
\includegraphics[width=1.0\hsize,angle=0]{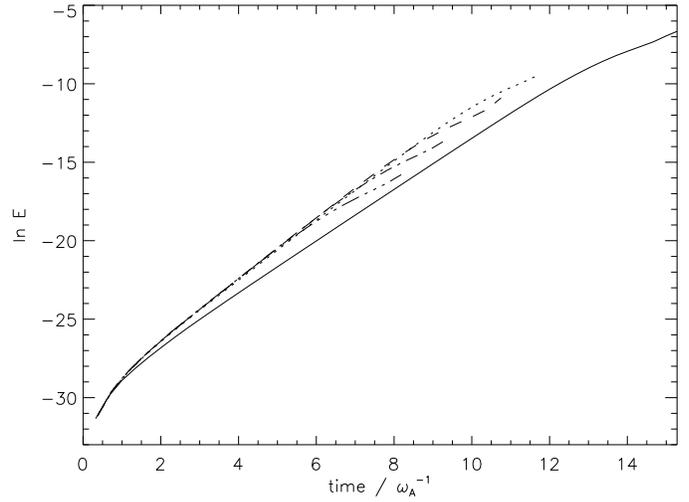}
\caption{{\mk The energy of the vertical component of
the magnetic field ($B_{\rm z}^2/8\pi$) against Alfv\'en time, for
five runs with different magnetic field strengths. The solid, dotted,
dashed, dot-dashed and dot-dot-dot-dashed lines represent runs with
$B_0=1.28, 0.404, 0.128, 0.0404, 0.0128$. The growth rate of the energy in the vertical component of
the magnetic field is roughly is roughly equal to
$2\omega_{\rm A}$. Adiabatic unstratified case.}}
\label{fig:bdep-051205}
\end{figure}

{\mk At this point, we can have a look at the growth of the poloidal field
component, since the production of a poloidal component from toroidal
component via this instability is part of the motivation for this
study. We can decompose the field into its three components (in
cylindrical coordinates). The strong-field run (with $B_0=1.28$,
$\omega_{\rm A}=0.81$) is used for this purpose, and the (log of the) energy
contained in the three components is plotted in
Fig.~\ref{fig:b_comps}, i.e. a volume integrations of $B_\phi^2/8\pi$,
$B_\varpi^2/8\pi$ and $B_z^2/8\pi$at $\varpi <
\varpi_{\rm max}=\varpi_0/2$, as in Eq.~\ref{eq:intovervarpi}. It can
be seen in the graph that the energy in the poloidal components
($B_\varpi$ and $B_z$) is growing exponentially, with a growth rate
double that of the growth in the displacement field -- owing to the
$B^2$ dependence.

In Fig.~\ref{fig:bdep-051205} is plotted the (log of the) energy in the $B_z$
component against the Alfv\'en time, i.e. $t/\omega_{\rm A}^{-1}$, for the
five runs of different field strength $B_0$ described above. It can be
seen that the growth rate of the poloidal field is indeed proportional
to $\omega_{\rm A}$, just as the growth rate in the displacement field
in Fig.~\ref{fig:bdep}. Note that the growth rate in the
strongest-field run is again a little lower than expected, for the reason
described above.}

\subsection{Rotation}
\label{sec:rot}

\begin{figure}
\includegraphics[width=1.0\hsize,angle=0]{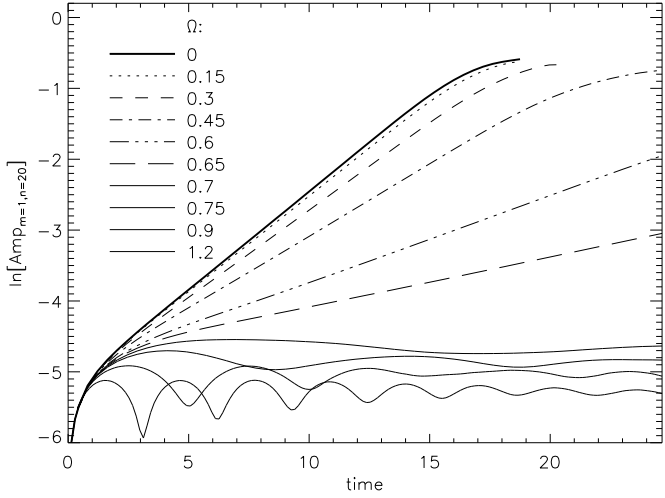}
\caption{Amplitudes of the $m=1$, $n=20$ mode at different values of
$\Omega$, taken from the displacement field. Adiabatic unstratified case.}
\label{fig:omega-disp}
\end{figure}

{\mk We shall now look at the effect of rotation on the
instability. An angular velocity is added (by means of a Coriolis
force), parallel to the magnetic axis and gravity.}
In the case of $p=1$ we expect, looking at (\ref{eq:mandprot}), to see the instability suppressed if
$\Omega\gg\omega_{\rm A}$. A number of non-stratified ($g=0$) runs were executed, using, as
before, a resolution of $72\times72\times8$, with $\eta=\kappa=0$. The values $0$,
$0.15$, $0.3$, $0.45$, $0.6$, $0.65$, $0.7$, $0.75$, $0.9$ and $1.2$ of
$\Omega$ were used, at a value of $B_0$ of $1.28$, which gives a
growth rate in the absence of rotation of $0.64$. In Fig.
\ref{fig:omega-disp} the amplitude of the $m=1$ mode (as taken from
the displacement field) is plotted
for each of the runs, as a function of time. It can be seen that the
instability is indeed suppressed if $\Omega$ is above a certain value
somewhere between $0.65$ and $0.70$. Above this value a distinct oscillatory
behaviour sets in, indicating stability.

In order to check that this result holds in the regime
$N\gg\omega_{\rm A}$, the runs were repeated using the stratified
setup (see Sect. \ref{sec:stratandnon}) and $g=9.1$, giving $N=5.8$
so that both $N\gg\omega_{\rm A}$ and $N\gg\Omega$ were fulfilled. This
was found to make no difference at all to the suppression of the
instability, there still existing a critical value of $\Omega$ roughly
equal to the non-rotating growth rate above which the instability was suppressed.

\subsection{Magnetic diffusion and its effect on high wavenumbers}
\label{sec:md}

We expect to to see an upper limit on $n$ due to magnetic
diffusion, which is the same as saying that we expect to see a maximum value of $\eta$ at which the
instability is seen, at a given value of $n$. As explained in Sect. \ref{sec:stratandnon}, we have to look
either at the growth of several $n$ modes at a given value of
diffusion $\eta$, or to look at the growth of
a particular $n$ mode in runs with different values of $\eta$. It was
found that the latter was easier to perform, hence we expect to see,
for a given value of $n$, a value of $\eta$ above which no instability is seen.
For these runs, the non-stratified setup was used as gravity has no
effect on this high-wavenumber limit.

We used $B_0=1.28$ as in the previous section, and the following
values of magnetic diffusivity $\eta$: $0, 2\times10^{-4},
4\times10^{-4}, 6\times10^{-4}, 10^{-3}, 1.6\times10^{-3},
2.4\times10^{-3}, 4\times10^{-3}, 6\times10^{-3}, 10^{-2}$ and
$1.6\times10^{-2}$.

\begin{figure}
\includegraphics[width=1.0\hsize,angle=0]{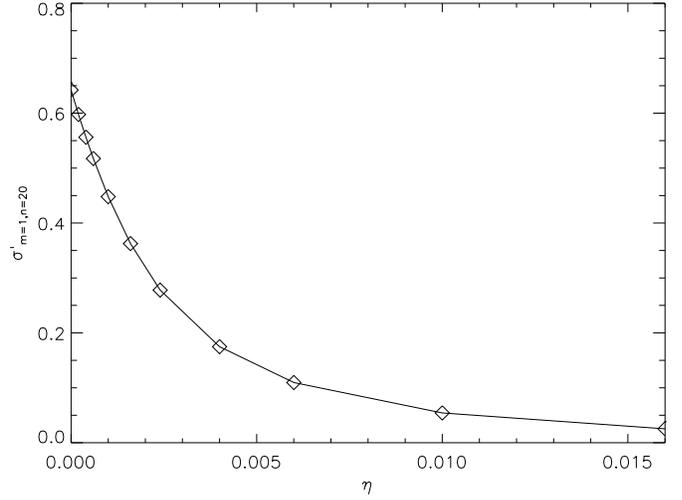}
\caption{Growth rate of the $m=1$, $n=20$ mode in the velocity field
as function of the magnetic diffusivity $\eta$. Unstratified case.}
\label{fig:diffusion}
\end{figure}

Measuring the growth of the mode at wavenumber $n=20$ as a function
of $\eta$, we expected to find a value of diffusion
$\eta_{\rm crit}\sim\sigma_0/n^2=1.6\times10^{-3}$ (from Eq.
(\ref{eq:minandmaxofn})) above which the instability
does not grow, where $\sigma_0$ is the growth rate in the absence of
diffusion. Fig. \ref{fig:diffusion} shows the growth rate of the
velocity field measured at the time at which the
growth rate in the zero diffusion (adiabatic) case is at its maximum
($t=10.4$). It can be seen in the figure that the instability is not
entirely suppressed at $\eta=\eta_{\rm crit}$, just slowed by a factor
of two or so. Even when $\eta$ is much higher than $\eta_{\rm crit}$
the growth rate is still above zero.

\subsection{Gravity and its effect on low wavenumbers}
\label{sec:grav}

We also expect to see a lower limit on the unstable vertical
wavelength $n$ determined by the value of $g$; $n_{\rm min} = N/\omega_{\rm A}
r$ (Eq. (\ref{eq:minandmaxofn})).

\begin{figure}
\includegraphics[width=1.0\hsize,angle=0]{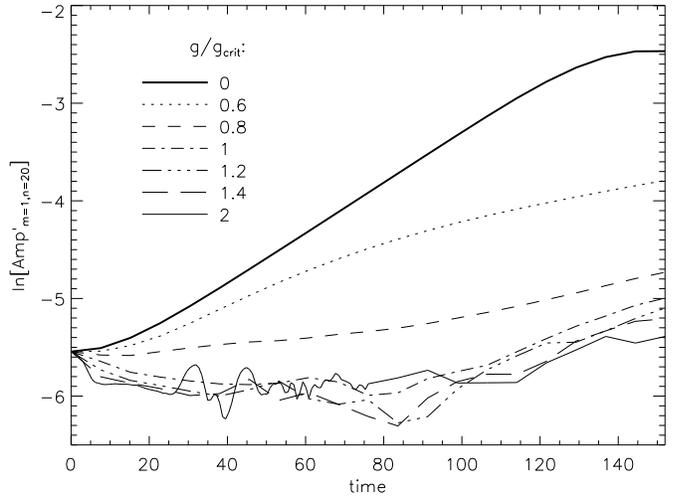}
\caption{Effect of the stabilising stratification, under adiabatic conditions
($\kappa=\eta=0$). Velocity field amplitudes of the $m=1$, $n=20$ mode as a
function of time for seven values of the acceleration of gravity $g$. The
first half of the run with $g=2g_{\rm crit}$ is plotted at higher time
resolution.}
\label{fig:gravitylimit}
\end{figure}

For these runs we obviously have to use the stratified setup, the
growth of the same mode, $n=20$, being measured as in the previous
three sections. All of the runs had $B_0=0.102$ and $\eta=\kappa=0$,
each run then having a different value of
gravitational acceleration $g$; the expected value of $g$ above
which the $n=20$ mode does not grow is $g_{\rm crit} = n B_0
\sqrt{5T/8\pi\rho} = 10/\pi$; the code was run with the following
multiples of this value: $0, 0.6, 0.8, 1, 1.2, 1.4$ and $2$. Fig.
\ref{fig:gravitylimit} shows the amplitude of the $m=1, n=20$ mode
(calculated from the velocity field) for these values of $g$. As the
figure shows, the instability stops growing at the expected value 
$g=g_{\rm crit}$, within the measurement accuracy.

\subsection{Thermal diffusion}
\label{sec:td}

\begin{figure}
\includegraphics[width=1.0\hsize,angle=0]{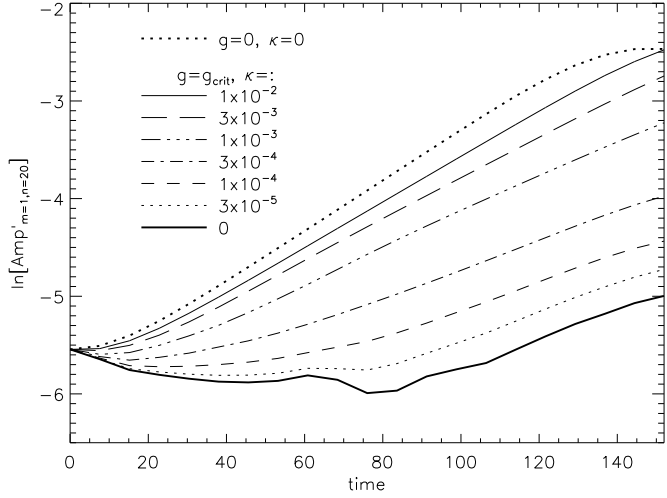}
\caption{Velocity field amplitudes of the $m=1$, $n=20$ mode as a
function of time, at different values of thermal diffusion $\kappa$, 
when $g=10/\pi$. For 
large values of $\kappa$, the stabilising effect of gravity is absent and
growth occurs at the same rate as in the adiabatic unstratified case 
(uppermost line).}
\label{fig:thermaldiffusion}
\end{figure}

If thermal diffusion is added, we expect to see a decrease in the minimum 
unstable 
vertical wavenumber, since the stable thermal buoyancy is reduced by diffusion
on small length scales. To investigate this, thermal diffusion is added
to the run in the previous section where $g=g_{\rm crit}=10/\pi$, i.e. the
borderline case where gravity is just strong enough to suppress the
instability. The values of $\kappa$ used were
$3\times10^{-5}$, $10^{-4}$, $3\times10^{-4}$, $10^{-3}$,
$3\times10^{-3}$ and $10^{-2}$. The results can be seen in
Fig. \ref{fig:thermaldiffusion}. The figure also shows, for
comparison, the case where $g=0$ and $\kappa=0$.

From (\ref{eq:minandmaxofn}) and (\ref{eq:thermdiff}), we have
\begin{equation}
n^2_{\rm min} = \frac{N^2}{\omega_{\rm A}^2r^2(1+\tau_{\rm I}/\tau_{\rm T})}
\propto \frac{g^2}{1+\kappa n_{\rm min}^2/\sigma}
\end{equation}

We expect, therefore, that an increase in both $g$ and $\kappa$ can
cancel each other out. It would be reasonable to expect that if we
increase $g$ from $0.8g_{\rm crit}$ to $g_{\rm crit}$, we need to increase
$\kappa$ from zero to $((1.0/0.8)^2-1)\sigma/n^2$ to cancel out the
effect on the instability. This works out as
$\kappa=9\times10^{-5}$. Likewise, from $g=0.6g_{\rm crit}$ we need
$\kappa=2.7\times10^{-4}$. This means that in runs with $g=g_{\rm crit}$,
using $\kappa=9\times10^{-5}$ and $2.7\times10^{-4}$ should produce the
same instability growth rate in the $m=1$, $n=20$ mode as if we had
reduced $g$ to $0.8g_{\rm crit}$ and $0.6g_{\rm crit}$ respectively, keeping $\kappa=0$.
As Fig. \ref{fig:thermaldiffusion} shows, these numbers are roughly in 
agreement with the numerical results.

\section{Discussion}

We set out in this study to verify, with numerical methods, the
results of the existing analytical work on the instabilities of
toroidal fields in stars, and to check that these
results were relevant, e.g. that a predicted instability is not
drowned out by stronger unstable modes
that might have been missed, for example due to simplifying
assumptions. The results have been largely positive.

Previous analytic work makes predictions of the dependence of
instability conditions and growth rates on the parameters of an azimuthal
magnetic field. These are the dependence of the field strength on distance 
$\varpi$ from the axis (the index $p={\rm d}\ln B/{\rm d}\ln\varpi$), the
stability of the stratification (measured by the buoyancy frequency $N$, or 
equivalently the acceleration of gravity $g$), the rotation rate $\Omega$,
the effects of magnetic diffusion
(diffusivity $\eta$), and the reduction of buoyancy by
thermal diffusion on small scales (diffusivity $\kappa$). 

First we checked the analytic prediction of which azimuthal orders $m$ should be
unstable, depending on the value of $p$ (Eq. (\ref{eq:mandp})). We
have confirmed this prediction (see Sect. \ref{sec:pandm}) for the cases
$p=1$ and $p=2$, the former being considered the most important as it
is this magnetic field which could plausibly be produced by the
winding-up of a seed field by differential rotation. The dominant mode
has $m=1$, a `kink' instability. 

The theory predicts that in the adiabatic, unstratified case ($\kappa=\eta=N=0$)
there is no threshold for instability, and that the growth rate $\sigma$ is of the 
order of the angular frequency $\omega_{\rm A}=r/v_{\rm A}$ of an Alfv\'en 
wave travelling around the star on an azimuthal field line. The field strength and 
hence $\omega_{\rm A}$ varies through the computational volume, but since the 
instability is a local one, the growth should be dominated by the largest value of 
$\omega_{\rm A}$ in the volume, after an initial transient. The numerical results 
reproduce this well. In the best studied case, for example, a value $\sigma/ 
\omega_{\rm A}=0.92$ was measured.

In the adiabatic case the prediction (Pitts \& Tayler 1986) is that the instability is 
suppressed when the rotation rate exceeds the Alfv\'en frequency $\omega_{\rm A}$. 
This was also verified, give or take a few percent at the most. The adiabatic case
is somewhat singular with respect to the effect of rotation, however. Theory
predicts that in the presence of strong thermal diffusion (${\kappa/\eta}
\rightarrow\infty$) the threshold for instability disappears again, and that the
growth rate is then of the order $\sigma=\omega^2_{\rm A}/\Omega$. We have
not tested this dependence, since the calculations required for this limiting case are 
computationally rather demanding.  

The effect of gravity and both magnetic and thermal diffusion were
investigated. The effect of gravity was
as expected -- above a certain value of $g$ the initial equilibrium is
stable at a given vertical wavenumber $n$. This translates into a
minimum unstable wavenumber which increases with increasing $g$.

The effect of magnetic diffusion on
the shorter wavelengths was not exactly the same as that expected. It
was found that at a given wavenumber $n$, a value of $\eta$ of the order suggested by
Acheson (1978) and Spruit (1999) did not kill the
instability entirely, rather it reduced its growth rate by about half. An
increase in $\eta$ beyond this reduced the growth rate still further,
but not to zero. Therefore it seems that magnetic diffusion alone
cannot suppress the instability.
It is possible that the discrepancy arises because the case $\eta\rightarrow
\infty$ may have a singular limit. The theoretical prediction was made for the 
case $\kappa=0$, but the relevant parameter is actually $\kappa/\eta$. By analogy 
with other double-diffusive systems, instabilities must exist also in the case 
$\kappa<\eta$ which do not appear when $\kappa=0$ is set from the
beginning. Since this case $\kappa<\eta$ is not of much astrophysical relevance, we
have not pursued this further.

Finally, the effect of thermal diffusion was tested. In the runs
executed, it was expected that values of $\kappa$ of the order of $9\times10^{-5}$
and $2.7\times10^{-4}$ would be needed to make a run with $g=g_{\rm crit}$
behave like the runs with $g=0.8g_{\rm crit}$ and $g=0.6g_{\rm crit}$
respectively. This appears to be correct, given that these were only
order-of-magnitude approximations.

The non-linear effect of the instability in all the above cases was found to be rather
simple. The net effect is similar to that of an enhanced magnetic 
diffusivity: the field configuration spreads horizontally, while the 
mean azimuthal magnetic flux decreases due to effective reconnection across the 
magnetic axis. Toroidal fields in stars are therefore predicted to
decay quickly by Tayler 
instability, once conditions for instability are satisfied, unless regenerated by 
differential rotation. {\mk Work continues on the non-linear evolution of
this instability, and some first results can be found in Braithwaite (2005).}


\end{document}